\documentclass[doublecol]{epl2} 

\title{Wave reflection in dielectric media obeying spatial Kramers-Kronig relations}
\shorttitle{Wave reflection in dielectric media ... } 

\author{S. Longhi \inst{1,2}}
\shortauthor{S. Longhi}

\institute{                    
  \inst{1}  Dipartimento di Fisica, Politecnico di Milano, Piazza L. da Vinci 32, I-20133 Milano, Italy\\
  \inst{2}  Istituto di Fotonica e Nanotecnlogie del Consiglio Nazionale delle Ricerche, sezione di Milano, Piazza L. da Vinci 32, I-20133 Milano, Italy}

\abstract{In a recent work, S.A.R. Horsley and coworkers [{\it Nature Photon.} {\bf 9}, 436-439 (2015)] 
showed rather interestingly that planar dielectric media, for which 
the real and imaginary parts of the dielectric permittivity are related by 
spatial Kramers-Kronig (KK) relations, are one-way reflectionless, whatever the angle of incidence.
Such a fascinating property, besides of extending our comprehension of the fundamental  phenomenon of reflection, may offer new ways for the design of antireflection surfaces and thin materials with efficient light absorption. However, KK dielectric media are generally described by slowly-decaying complex permittivity profiles which may introduce some subtle issues in the study of the scattering problem.  Here we provide a condition on the imaginary part of the dielectric profile that ensures the existence of proper scattering states and present a proof of the one-way reflectionless property in KK dielectric media which is free from loophole. Finally, we show that instabilities might arise at the interface of KK dielectric media when the medium is not purely dissipative.}
\begin{document}

\maketitle

\section{1. Introduction}
Kramers-Kronig (KK) relations are well known in optics to relate the real and imaginary parts of the complex  dielectric function of a medium in the frequency domain \cite{r1}. In a recent work, Horsley {\it et al.} introduced the idea of {\it spatial} KK relations  in planar  inhomogeneous dielectric media \cite{r2} (see also Ref.\cite{r2bis}). A remarkably property of such media is that, whenever the spatial distributions of real and imaginary parts of the dielectric permittivity are related by KK relations, they will not reflect waves incident from one side, whatever the angle of incidence. One-way reflectionless dielectric media of KK type are thus described by a complex permittivity  $\epsilon(x)$ that can be prolonged in the complex $x$ plane and that is analytic and bounded in the upper (or lower) half complex plane. Roughly speaking, the lack of back scattering in a KK medium stems from the absence of negative (or positive) spatial Fourier components of $\epsilon(x)$ \cite{r2}, and can be regarded as a generalization of the one-way reflectionless property found in locally-periodic $\mathcal{PT}$-symmetric optical media where Bragg scattering acts in one direction solely \cite{r3,r4,r5,r6,r7,r8}.There are, however, some subtle issues that the authors of Ref.\cite{r2} did not consider in their analysis. \\ The first issue arises because locally inhomogeneous dielectric profiles $\epsilon(x)$ obeying spatial KK relations are slowly-decaying potentials. While scattering of waves in short-range potentials (such as those satisfying Faddev's condition \cite{r12}) is a well-established problem, slowly-decaying potentials have been only quite recently studied in the mathematical literature and they may show unusual properties in terms of spectrum and existence of bound states embedded in the continuum (see, for instance, \cite{r9bis,r10,r11}). The scattering problem in slowly-decaying potential should be formulated by properly considering WKB asymptotic solutions of the wave equation, which are not in the most general case plane waves (Jost functions) \cite{r10,r11}. For KK permittivity profiles, we show that the scattering problem is well-posed provided that a minimal condition on the imaginary part of the dielectric permittivity is satisfied. The second issue is a loophole in the proof of the one-way reflectionless property of KK media given in Ref.\cite{r2}. In \cite{r2} a Born series expansion  method is used that assumes {\it a priori} plane waves as asymptotic solutions (see Eqs.(4)and (5) of Ref.\cite{r2}), however such an assumption fails in a large class of KK media that violate the so-called 'cancellation condition' discussed in this paper. Also, there is not any proof about the convergence of the Born series.  A way to overcome such an inconsistency has been suggested in Ref.\cite{r2bis}   using WKB and Stokes analysis. Here we re-formulate the scattering problem in a form which is consistent with the slow decay nature of KK potentials \cite{r10,r11} and provide an alternative proof of the one-way reflectionless property which is free from loophole. The third issue is related to dielectric media obeying KK relations which are not purely dissipative. In the presence of gain, unstable growing waves can arise at the interface separating the two asymptotically homogeneous regions of the medium. 

\section{2. Wave reflection in a planar inhomogeneous dielectric medium with slowly decaying profile}
\subsection{Wave equation and Kramers-Kronig dielectric media}
Let us consider the electromagnetic scattering problem of a monochromatic optical wave at frequency $\omega$ across an inhomogeneous planar dielectric medium in the $(x,y)$ plane. We focus our analysis to TE-polarized optical waves, with the electric field $\mathcal{E}$ vibrating in the direction $z$ orthogonal to the $(x,y)$ plane and the magnetic vector $\mathcal{H}$ lying in the $(x,y)$ plane . After setting $\mathcal{E}(x,z,t)=\mathbf{E}(x,y) \exp(-i \omega t)+c.c.$, $\mathcal{H}(x,z,t)=\mathbf{H}(x,y) \exp(-i \omega t)+c.c.$, for TE-polarized waves ($E_x=E_y=H_z=0$), the electric field component $E_z(x,y)$ satisfies the Helmholtz equation
\begin{equation}
\frac{\partial^2 E_z}{\partial x^2}+ \frac{\partial^2 E_z}{ \partial y^2}+k_0^2 \epsilon(x) E_z=0
\end{equation}
where $k_0= \omega / c_0$ and $\epsilon(x)$ is the relative dielectric permittivity, which is assumed of the form (using the same notations as in \cite{r2})
\begin{equation}
\epsilon(x)=\epsilon_b+ \alpha(x).
\end{equation}
In Eq.(2), $\epsilon_b$ is the substrate permittivity (real and positive number), and $\alpha(x)$ vanishes as $x \rightarrow \pm \infty$, so that to  describe a localized inhomogeneity near $x=0$. $\alpha(x)$ is generally assumed to be complex, i.e. the medium in the inhomogeneous region can provide local loss and/or gain. Following Ref.\cite{r2}, a KK dielectric medium which is reflectionelss for left incidence side corresponds to a function $\alpha(x)$ which is the boundary value on the real $x$ axis of an analytic and bounded function in the upper half complex plane. In such a case the real and imaginary parts of $\alpha(x)$ are related by a Hilbert transform (spatial KK relations), namely
\begin{eqnarray}
{\rm Re} [\alpha(x)] & = & \frac{1}{\pi} {\rm P} \int_{-\infty}^{\infty} \frac{\rm{Im}[\alpha(s)]}{s-x} ds \\
{\rm Im} [\alpha(x)] &= & -\frac{1}{\pi} {\rm P} \int_{-\infty}^{\infty} \frac{\rm{Re}[\alpha(s)]}{s-x} ds
\end{eqnarray}
where $\rm{P}$ denotes the principal value.
\subsection{Scattering problem: reflection and transmission coefficients}
 We look for a solution to Eq.(1) corresponding to an optical wave incident onto the inhomogeneous region of the dielectric from the left side (i.e. coming from $x\rightarrow - \infty$) at some incidence angle $\theta$ ($\theta \rightarrow 0$ for grazing incidence), see Fig.1(a). Such a solution has the form
 \begin{equation}
 E_z(x,y)=e_z(x) \exp(ik_yx)
 \end{equation}
where $0<k_y< k_0 \sqrt{\epsilon_b}$. The incidence angle $\theta$ is given by 
$ \theta={\rm atan} (k_x / k_y)$, where we have set $k_x=\sqrt{k_0^2 \epsilon_b-k_y^2}>0$. The amplitude $e_z(x)$ then satisfies the stationary one-dimensional Schr\"{o}dinger-like wave equation
\begin{equation}
E e_z(x)= \hat{H} e_z(x)
\end{equation}
where $E=k_x^2=k_0^2 \epsilon_b-k_y^2$ and $\hat{H}=-(d^2 /dx^2)+V(x)$ is the Hamiltonian with the optical potential $V(x)$ given by
\begin{equation}
V(x)=-k_0^2 \alpha(x)=-k_0^2 [\epsilon(x)-\epsilon_b].
\end{equation}
When the potential has a limited support or is short-range, i.e. it decays at infinity sufficiently fast, the scattering states to Eq.(6) have the plane-wave  asymptotic form $e_z(x) \sim \exp( \pm i k_x)$ as $x \rightarrow \pm \infty$ (Jost functions). A sufficient condition for the existence of Jost functions is provided by Faddev's condition \cite{r12}. However, for dielectric media satisfying spatial KK relations (3) and (4) the resulting optical potential $V(x)$ is generally long-range. In fact, let us indicate by $V_R(x)=-k_0^2 {\rm Re}(\alpha(x))$ and $V_I(x)=-k_0^2 {\rm Im}(\alpha(x))$ the real and imaginary parts of the optical potential, respectively. Then, owing to a general theorem by Kober \cite{r13} if $V_I(x)$ belongs to $L^1$, i.e. if
$\int_{-\infty}^{\infty} dx |V_I(x)| < \infty$, then in general $V_R(x)$ does not belong to $L^1$. The same holds by reversing $V_R$ and $V_I$. A necessary condition for {\it both} $V_R$ and $V_I$ to belong to $L^1$ is given by the {\it cancellation condition} \cite{r13}
\begin{equation}
\int_{- \infty}^{\infty} dx V(x) =0.
\end{equation}
 For example, in purely dissipative media,  $V_I(x) \leq 0$, $V(x)$ does not belong to $L^1$ because the cancellation law is not satisfied. Hence, for the most general class of KK dielectric media the scattering states are not of Jost (plane-wave) type, rather they are of the WKB type \cite{r10,r11}
\begin{equation}
e_z(x) \sim \exp \left[ \pm i k_x x \mp \frac{i}{2k_x} \int_0^x d \xi V( \xi) \right]
\end{equation}
 and the limits $\lim_{x \rightarrow \pm \infty} \int_0^x d \xi V(\xi)$ do not necessarily are finite.
 While the authors of Ref.\cite{r2} recognized such a circumstance when considering a special analytically-solvable model, their general proof  that KK media are one-way reflectionelss shows a loophole. In fact, the proof uses a standard Born series expansion method which assumes Jost functions as asymptotic states, however  for KK optical media that violate the cancellation condition mentioned above such an {\it a priori} assumption is incorrect. The scattering problem should be reformulated in a consistent way, which  properly accounts for the slow-decay nature of KK optics potentials and uses the WKB asymptotic states (9) for the incoming and scattered waves.  To study wave scattering in slowly-decaying potentials, we follow the general procedure outlined in Refs.\cite{r10,r11} and define the {\it generalized} transmission $t$ and reflection $r$ coefficients (for left-side incidence) from the solution to Eq.(6) with the asymptotic behavior
 \begin{equation}
 e_z(x) = \left\{ 
 \begin{array}{cc}
 \exp( i \phi)+r(k_x) \exp(-i \phi) +o(1) & x \rightarrow - \infty \\
 t(k_x) \exp(i \phi) +o(1) & x \rightarrow  \infty
 \end{array}
 \right .
 \end{equation}
  where we have set
  \begin{equation}
  \phi=\phi(x,k_x)=k_xx-(1/2k_x) \int_0^x d \xi V(\xi).
  \end{equation}
  If the limits $\lim_{x \rightarrow \pm \infty} \int_0^x d \xi V(\xi)$ exist and are finite, then $\phi(x,k_x)$ can be replaced by $k_xx$  in Eq.(11) \cite{r11}, the usual plane-wave asymptotic states are found and $t,r$ give ordinary transmission and reflection coefficients after suitable renormalization of amplitudes involving  the integrals $\exp [ i /(2k_x) \int_{0}^{\pm \infty} V(x) dx ]$.  However, if $\lim_{x \rightarrow \pm \infty} \int_0^x d \xi V(\xi)= \infty$, even though the potential $V(x)$ is weak the substitution $\phi = k_x x$ is not permitted since $ \int_0^x d \xi V(\xi)$ secularly grows as $x \rightarrow \pm \infty$. In case  $\lim_{x \rightarrow \pm \infty} \int_0^x d \xi V(\xi)= \infty$, the scattering problem (10) is well-posed provided that the WKB solutions $\exp[ \pm i \phi(x)]$ do not secularly grow (or decay) as $|x| \rightarrow \infty$. While this is always the case for a real-valued potential $V(x)$ \cite{note2}, for a complex dielectric permittivity (and hence for a complex potential $V$) the WKB solutions $\exp[ \pm i \phi(x)]$ do not secularly grow nor decay at infinity provided that the limits
 \begin{equation}
 \lim_{x \rightarrow \pm \infty} \int_{0}^{x} d \xi {\rm Im} [\alpha(\xi)]
 \end{equation}
  are finite. In KK media where such a condition is violated, the scattered states do not have an asymptotic constant intensity, and this makes the scattering problem and definition of reflection/transmission coefficients questionable. For example, let us consider the dielectric medium corresponding to a single pole of $\epsilon(x)$ in the lower half plane, i.e.
 $\alpha(x)= \epsilon(x)-\epsilon_b=A/(x+i \xi)$
  with $\xi>0$ and $A$ an arbitrary constant. In Ref.\cite{r2} the authors limited themselves to consider the case $A$ real and negative. However, Eq.(13) describes a KK dielectric medium for any arbitrary value of the constant $A$, which can also take complex values. For such a specific example one obtains from Eq.(11)
\begin{equation}
\phi(x)=k_x x-[A/(2k_x)] {\rm ln} \left( 1-ix/ \xi \right)
\end{equation} 
 At large values $|x|$,  apart from an inessential constant phase term the asymptotic behavior of $\phi(x)$ is given by
 $ \phi(x) \sim k_x x -[A/(2 k_x)] {\rm ln} |x|$
so that $| \exp[ \pm i \phi(x)] | \sim |x|^{\mp A_I /(2k_x)}$ as $|x| \rightarrow \infty$, where $A_I$ is the imaginary part of $A$. Hence when $A_I \neq 0$ the WKB states (9)  show secularly growing or damped amplitudes, and a serious issue arises in the definition of the scattering problem and reflection/transmission coefficients. For example, if $\exp[i \phi(x)] \rightarrow 0$ (and thus $\exp[-i \phi(x)] \rightarrow \infty$)  as $x \rightarrow -\infty$, even if $r \rightarrow 0$ in Eq.(10) there could be a reflected wave because $r$ is multiplied by the WKB wave $\exp[-i \phi(x)]$ ]with diverging intensity. 

\subsection{Grazing incidence}
The scattering problem discussed in the previous subsection can be re-formulated in terms of a scattering problem for the non-stationary Schr\"{o}dinger equation when considering grazing incidence, i.e. for a small incidence angle $\theta$ (see for example \cite{r14}). For grazing incidence, we look for a solution to the wave equation (1) of the form
\begin{equation}
E_z(x,y)=\psi(x,y) \exp(ik_0 \sqrt{\epsilon_b} y)
\end{equation}
 where $\psi(x,y)$ is a slowly-varying function with respect to the variable $y$ as compared to the exponential term.  By neglecting the second derivative of $\psi(x,y)$ with respect to $y$ (like in standard paraxial approximation), substitution of Eq.(14) into Eq.(1) yields the non-stationary Schr\"{o}dinger-like wave equation
 \begin{equation}
 2 i k_0 \sqrt{\epsilon_b} \frac{\partial \psi}{\partial y}= -\frac{\partial^2 \psi}{\partial x^2}+V(x) \psi \equiv \hat{H} \psi
 \end{equation}
 where the optical potential is again given by Eq.(7). The scattering problem in terms of wave packets, which is suited for a numerical analysis and that would correspond to an experimental setting to measure reflection and transmission at the interface, can be now formulated as follows: for a given initial wave packet distribution  $\psi(x,0)$ at $y=0$, which is localized far from the inhomogeneous region on the left side ($x \rightarrow - \infty$), the wave packet is propagated along the $y$ direction according to $\psi(x,y)= \exp[-i \hat{H} y /(2 k_0 \sqrt{\epsilon_b})] \psi(x,0)$, and the final (asymptotic) state after the scattering process is determined in the $ y \rightarrow + \infty$ limit of the solution $\psi(x,y)$. Interestingly, such a formulation of the scattering process in terms of finite wave packets   is well-posed even for KK permittivity profiles  that violate the cancellation condition (12), and can be suitably used to check the appearance of unstable modes at the interface, as shown in a following section.

 \section{3. One-way reflectionless property of Kramers-Kronig dielectric media}
 A proof of the reflectionless property of KK media which is free from loopholes can be obtained by formulating the scattering problem as in Ref.\cite{r11}, i.e. by the introduction of the envelopes $w_1(x)$ and $w_2(x)$ via the Anstaz
 \begin{eqnarray}
 e_z(x) & = & w_1(x) \exp(i \phi )+w_2(x) \exp(-i \phi) \\
 \frac{ d e_z(x)}{dx} & = & i k_x \left[ w_1(x) \exp(i \phi )-w_2(x) \exp(-i \phi) \right].
 \end{eqnarray}
 The envelopes $w_{1,2}$ satisfy the coupled equations
 \begin{eqnarray}
 \frac{dw_1}{dx} & = & -\frac{iV(x)}{2k_x} w_2(x)  \exp(-2i \phi )\\
\frac{dw_2}{dx} & = & \frac{iV(x)}{2k_x} w_1(x)  \exp(2i \phi ).
 \end{eqnarray} 
 Equations (18) and (19) should be integrated with the boundary conditions 
 \begin{eqnarray}
 w_2(x) \rightarrow 0 \; \; {\rm as} \;\; x \rightarrow + \infty \\
 w_1(x) \rightarrow 1  \; \; {\rm as} \;\; x \rightarrow - \infty, 
 \end{eqnarray}
  corresponding to a wave incident onto the interface region from the left side. The generalized reflection and transmission coefficients are then given by
 \begin{eqnarray}
  r(k_x) =\lim_{x \rightarrow - \infty}w_2(x) \\
  t(k_x) =\lim_{x \rightarrow  + \infty}w_1(x).
  \end{eqnarray}
   Integrating formally both sides of Eq.(19) from $-\infty$ to $\infty$ and using Eqs.(20) and (22), one can write 
   \begin{equation}
   r(k_x)=-\frac{i}{2k_x} \int_{-\infty}^{\infty} dx w_1(x)V(x) \exp[2i \phi(x)].
   \end{equation}
   Note that, as compared to more conventional WKB analysis of Ref.\cite{r2bis}, following Ref.\cite{r11} we use for the asymptotic states the expression (9) , which offers the advantage of avoiding branch cuts arising form square root of the potential and simplifies the analysis when extending the equations into the complex plane. Since $V(x)$, $\phi(x)$ and $w_1(x)$  can be continued into the complex plane $z$ (the real axis is ${x=\rm Re}(z)$), the integral on the right hand side in Eq.(24) can be computed by deforming the contour in complex plane. To this aim, let us first notice that,  since $V(z)$ is analytic in the half-plane ${\rm Im}(z) \geq 0$, the phase $\phi(x)$ can be extended in the upper half complex plane by assuming 
 \begin{equation}
 \phi(z)=k_xz+\int_0^{z}  d \xi V(\xi),
 \end{equation}
 where the integral is taken along an arbitrary path $\gamma$ in the upper half plane that connects the origin $\xi=0$ with $\xi=z$ [see Fig.1(b)]. Since $V(z)$ is analytic in the half-plane ${\rm Im}(z) \geq 0$, the integral on the right hand side in Eq.(25) is independent of the path $\gamma$ and thus $\phi(z)$ is holomorphic in the domain ${\rm Im} (z) \geq 0$. By the replacement $x \rightarrow z$ in Eqs.(18) and (19),  analytic continuations of $w_1$ and $w_2$ can be obtained as well, which are holomorphic in the ${\rm Im}(z) \geq 0$ domain because the matrix $\mathcal{M}(z)$ of the differential system defines a holomorphic map in such a half plane.   
 After setting
 \begin{equation}
 f(z)=\frac{1}{2 i k_x} V(z) w_1(z)  \exp[2i \phi(z)]
 \end{equation} 
 since $f(z)$ is analytic in the domain ${\rm Im}(z) \geq 0$, one has $ \oint_{\sigma} f(z)dz=0$ along the closed contour $\sigma=\sigma_1 \bigcup \sigma_R$ shown in Fig.1(b). Taking the limit $R \rightarrow \infty$ and using Eq.(24) one obtains
 \begin{eqnarray}
 r(k_x) & = &  \lim_{R \rightarrow \infty} \int_{\sigma_R} f(z) dz \nonumber \\
 &= &  \lim_{R \rightarrow \infty} iR \int_0^{\pi} d \theta \exp(i \theta) f(R \exp(i \theta)).
 \end{eqnarray}
To calculate the integral on the right hand side of Eq.(27), it is worth determining upper bounds for $V(z)$, $w_1(z)$ and $\exp[2 i \phi(z)]$ along the semi-circle $z=R \exp(i \theta)$  ($ 0 \leq \theta \leq \pi$) as $R \rightarrow \infty$. To this aim, let us assume that $V(z) \rightarrow  0$ as $|z| \rightarrow \infty$ like $V(z) \sim C/z^{ \delta}$ with $\delta \geq 1$, where the constant $C \neq 0$ is real-valued when $\delta=1$ [this is to satisfy the constraint (12)], whereas it can assume complex values when $\delta>1$ \cite{noteuff}. Therefore on the semi-circle $z=R \exp (i \theta)$ for $R \rightarrow \infty$ one has $|V(z) \exp[2i \phi(z)]| \leq C' \exp(-2k_x R \sin \theta) /R^{\delta}$ for some constant $C'>0$. The asymptotic behavior of $w_1(z)$ as $|z| \rightarrow \infty$ can be readily determined after elimination of $w_2$ in Eqs.(18) and (19). One obtains
 \begin{equation}
\frac{d^2 w_1}{dz^2}-\left( \frac{dV/dz}{V} - 2 i k_x +\frac{i V}{k_x} \right) \frac{dw_1}{dz}-\left( \frac{V}{2k_x}\right)^2 w_1=0.
 \end{equation}
 From the asymptotic behavior of the coefficients in Eq.(28) as $|z| \rightarrow \infty$, it can be readily shown that there are two linearly-independent solutions $\varphi_1(z)$ and $\varphi_2(z)$ to Eq.(28) with the  asymptotic behavior $\varphi_1 (z) \sim 1$ and $\varphi_2(z) \sim \exp[-2i \phi(z)]$  as $|z| \rightarrow \infty$. The most general solution to Eq.(28) is given by a linear superposition of $\varphi_1(z)$ and $\varphi_2(z)$, namely $w_1(z)= \alpha_1 \varphi_1(z)+ \alpha_2 \varphi_2(z)$. The coefficients $\alpha_{1,2}$ of the linear combination can be determined by observing that, for $z=x$ real, $w_1(x) \sim 1$ as $x \rightarrow -\infty$ [see Eq.(23)]. This yields $\alpha_1=1$ and $\alpha_2=0$, i.e. $w_1(z)= \varphi_1(z)$. Since $\varphi_1(z) \sim 1 $ as $|z| \rightarrow \infty$, $w_1(z)$ is a limited function of $z$. Hence over the semi-circle $z=R \exp(i \theta)$ one has $|f(z)| \leq  (\kappa / k_x)  \exp(-2k_x R \sin \theta) / R^{\delta}$ for some constant $\kappa$, and thus 
\begin{eqnarray}
\left| \int_{\sigma_R} f(z) dz \right| \leq R \int_{0}^{\pi} d \theta \left| f(R \exp(i \theta) ) \right| \nonumber \\
\leq \frac{\kappa}{ k_x R^{\delta-1}}  \int_0^{\pi} d \theta \exp(-2k_x R \sin \theta) \leq \\
\frac{ 2 \kappa}{k_x R^{\delta-1}}  \int_0^{\pi /2} d \theta \exp(-4k_x R \theta / \pi) 
\leq \frac{\pi \kappa}{2 k_x^2 R^{\delta}}
\rightarrow 0 \nonumber
\end{eqnarray}
as $R \rightarrow \infty$. From Eqs.(27) and (29) it follows that $r(k_x)=0$ for any $k_x>0$, i.e. for an arbitrary incidence angle. Note also that from the above analysis one has $w_1(z)= \varphi_1(z) \rightarrow 1$ as $|z| \rightarrow \infty$. Hence from Eq.(23) one obtains 
\begin{equation}
t(k_x)=1.
\end{equation}
 Note that, since $t$ is a generalized transmission coefficient, the condition  (30) does not imply that the intensity of the transmitted wave is the same than the one of the incident wave. The transmittance $T$ can be readily obtained from Eqs.(10) and (30) taking into account the asymptotic behavior of the modulus square of the WKB states $\exp[ \pm i \phi(x)]$ as $x \rightarrow \pm \infty$. One obtains
 \begin{equation}
 T=\exp \left[ \frac{1}{k_x} \int_{-\infty}^{\infty} {\rm Im}(V(x)) dx \right].
 \end{equation}
Note that for a medium with balanced gain and loss one has $T=1$. 

 \section{3. Unstable modes at the interface}
 Another subtle issue which was not considered in the analysis of Ref.\cite{r2} is the appearance of possible instabilities near the interface $x=0$ of the inhomogeneous dielectric medium. Such an instability can be at best explained in case of grazing incidence, for which the scattering problem can be formulated in terms of the non-stationary Sch\"{o}dinger equation (15): an instability arises when an initial normalizable field distribution $\psi(x,0)$ (wave packet) at $y=0$ shows a secular growth near the interface $x=0$ as $y \rightarrow \infty$. Such an instability obviously requires some gain in the medium, i.e. it is always absent in purely dissipative media. The permittivity profiles presented in the copious examples of Ref.\cite{r2} seem free from instabilities, however one can find KK profiles where instabilities can arise.  
 Since the Hamiltonian $\hat{H}$ in the Schr\"{o}dinger equation (17) is non-Hermitian, two different kinds of instabilities can arise: (i) from bound states of $\hat{H}$ with complex energy $E$ (i.e. belonging to the point spectrum of $\hat{H}$), corresponding to an exponential growth of $\psi(x,y)$ with $y$ and an imaginary $k_y$ wave number;  (ii) from exceptional points in the continuum, corresponding to an algebraic growth of $\psi(x,y)$ with $y$ \cite{r15}. Here we briefly discuss the latter case, which is amenable to analytical treatment. Let us consider the following dielectric permittivity profile $\alpha(x)=A/(x+i \xi)^2$, corresponding to the optical potential
 \begin{equation}
 V(x)=-Ak_0^2/(x+i \xi)^2
  \end{equation}
 with $\xi>0$. Note that $V(x)$ has a second-order pole at $x=-i \xi$ and it is $\mathcal{PT}$-symmetric. Since $\int_{-\infty}^{\infty} dx V(x)$ is finite, the WKB asymptotic form of scattering states $\exp[\pm i \phi(x)]$  are plane waves (Jost functions), and the potential is reflectionless for left incidence side whatever the value of the amplitude $A$. The dielectric profile (32) was  considered in Ref.\cite{r2} for $A>0$, and instabilities were not observed in numerical simulations. Here we focus our attention to the case $A<0$.   For $A<0$, the problem is exactly solvable by means of a cascade of supersymmetric (SUSY) transformations when the amplitude $A$ takes the values   $A= A_n=-n(n+1)/k_0^2$ ($n=1,2,3,...$) (see  \cite{r15}). At such amplitude values, the non-Hermitiain Hamiltonian $\hat{H}$ has an exceptional point of order $(n-1)$ at $E=0$, i.e. at the bottom edge of the continuous spectrum \cite{r15}. Let us assume, for the sake of definiteness, the lowest value of the amplitude $A$ admitting a first-order exceptional point ($n=2$), i.e. $A=A_2=- 6/k_0^2$ and $V(x)=6/{(x+i \xi)^{2}}$.
The exact scattering states of Eq.(6) can be calculated by a double SUSY and read explicitly
\begin{equation}
e_z(x)= \left(-k_x^2-\frac{3ik_x}{x+i \xi}  +\frac{3}{(x+i \xi)^2} \right) \exp(ik_x x)
\end{equation}
with energy $E=k_x^2$. Note that, like for the case $A>0$ considered in Ref. \cite{r2}, such a potential is reflectionless and even invisible (i.e. $t=1$) for both incidence sides. However, the edge $E=0$ of the continuous spectrum of $\hat{H}$ is an exceptional point of first order \cite{r15}, sustaining the bound state
\begin{equation}
\phi_b(x)=\frac{1}{(x+i \xi)^2}
\end{equation}
with associated function $\theta_a(x)=1/6$, i.e.
\begin{equation}
\hat{H} \phi_b=0 \; \;,\;\;\; \hat{H} \theta_a=\phi
_b\end{equation}
as one can check directly. The instability of the scattering states (33) can be proven as follows. Let $ \eta$ be an arbitrarily small parameter.  Then it can be readily shown that an exact solution to Eq.(1) with the initial condition $E_z(x,y=0)= \eta \theta_a(x)$ is given by 
\begin{equation}
E_z(x,y)=\eta \left(\theta_a(x)+ \frac{y}{2i k_0 \sqrt{\epsilon_b}} \phi_b(x) \right) \exp(i k_0 \sqrt{\epsilon_b} y)
\end{equation}
 which secularly grows with an algebraic law as $y \rightarrow \infty$. This means that initial small noise 'shaped' like the associated function $\theta_a(x)$ can trigger the secular growth of the bound mode $\phi_b(x)$ at the interface. Note that, even though the bound state $\phi_b(x)$ does not radiate along the $x$ direction, owing to its secular growth an observer placed on the left side of the interface at a distance $d$ and after a propagation length $y=y_0$ will detect a radiation of amplitude $|E_z| \sim \eta y_0/(k_0 \sqrt{\epsilon_b} d^2)$, which can not be negligible for a sufficiently large propagation distance $y_0$. The growth is even faster, with a power law $ \sim y_0^m$, for an exceptional point of order $m$ larger than one. 
 Beam propagation at grazing incidence in a dielectric medium with permittivity profile given by Eq.(32), as obtained by numerical solutions of the Schr\"{o}dinger equation (17) using a standard pseudo-spectral split-step method, is shown in Fig.2 for $A>0$ [panel (a)] and $A<0$ [panel (b)]. Parameter values used in the simulations are $\lambda=2 \pi /k_0=1$, $\xi=0.5 \lambda$, $\epsilon_b=2.25$ and $A= \pm 20/(k_0)^2= \pm 5 / \pi^2$. The initial condition is the Gaussian beam distribution $\psi(x,0)=\exp[-(x+q)/w^2-ik_x x]$ with spot size $w=25 \lambda$, offset $q=65 \lambda$ and $k_x=0.2/ \lambda$ (corresponding to an incidence angle $\theta \simeq 1.216^{\rm o}$). In both cases,  one observes light amplification near the interface region. However, while for $A>0$ amplified light is observed when the beam reaches the inhomogeneous region and it is convected away in the $x>0$ region, for $A<0$ one  clearly observes a rapid and secularly growing bound state at the interface near $x=0$. To highlight the different behavior in the two cases, Fig.2(c) depicts the evolution of the light intensity $I=|\psi(x=-d,y)|^2$ versus $y$ measured by an observer at the distance $d=30 \lambda$ on the left side of the interface: note that, while for $A>0$ after the passage of the incident beam the intensity $I$ decays to zero, for $A<0$ it increases again after the passage of the incident beam, which is the signature of the unstable growing mode. In a realistic medium, the growth of the unstable mode is obviously saturated by some nonlinear process. Nevertheless, nonlinear saturation would modify the effective permittivity profile for linear waves as well, breaking the exact reflectionless property of the interface. Such a brief analysis, far from being exhaustive, shows that the issue of instabilities in not purely dissipative KK media should deserve further consideration.

 \begin{figure}
\onefigure[width=8.8cm]{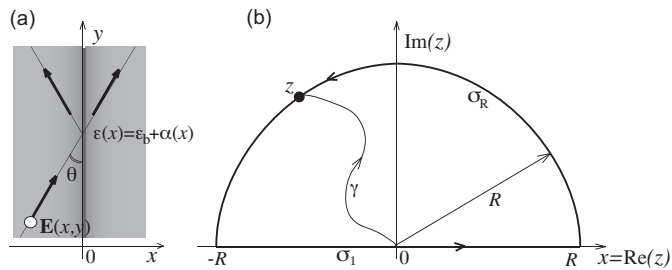}
\caption{(Color online) (a) Scattering of TE-polarized waves in an inhomogeneous dielectric medium. (b) Contour path $\sigma= \sigma_1 \cup \sigma_R$ in complex $z$ plane used to calculate the reflection coefficient $r(k_x)$.}
\end{figure}

\begin{figure}
\onefigure[width=8.8cm]{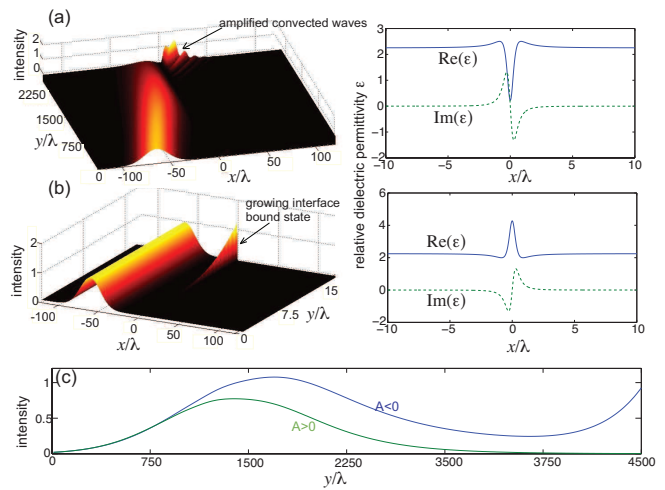}
\caption{(Color online) Scattering of a Gaussian beam at grazing incidence in the dielectric medium with permittivity $\epsilon(x)=\epsilon_b+A/(x+ i \xi)^2$ for $\lambda=2 \pi / k_0=1$, $\xi=0.5 \lambda$,  $\epsilon_b=2.25$  and for (a) $A=5/ \pi^2$, and (b) $A=-5 / \pi^2$.  Other parameter values are given in the text. The left panels show the numerically-computed evolution of the beam intensity $|\psi(x,y)|^2$, whereas the right panels show dielectric permittivity profiles $\epsilon(x)$. Beam propagation is shown for a limited propagation distance $y$, above which large wave amplification arise. (c) Numerically-computed light intensity as a function of $y/ \lambda$ at the position $x=-d=-30 \lambda$.}
\end{figure}

\section{4. Conclusions} 
In a recent Letter \cite{r2}, Horsley {\it et al.} discovered the fascinating property that optical dielectric interfaces with  spatial permittivity profiles whose analytic continuation are holomorphic functions in the upper (or lower) half part of the complex plane are one-way reflectionless.
 However, KK dielectric media are generally described by slowly-decaying permittivity profiles which may introduce some subtle issues. Here we re-considered the analysis of Ref.\cite{r2} and suggested a different route to prove the one-way reflectionless property of KK dielectric media, which is free from loophole, and pointed out that an additional constraint should be imposed to properly formulate the scattering problem. We also briefly mentioned the existence of unstable states at the interface of KK dielectric media when the medium is not purely dissipative, an issue that deserves further analysis. The present results are expected to motivate further investigations on the scattering properties of KK dielectric media and, more generally, of optical media with slowly decaying potentials, where unusual properties (such as the appearance of infinitely many bound states embedded in the continuum \cite{r9bis}) could be observable in an optical setting.
\acknowledgments The author acknowledges fruitful discussions with A. Mostafazadeh and S.A.R. Horsley.


\begin{thebibliography}{0}
\bibitem{noteuff}
The proof that $r(k_x)=0$ given here holds in the more general case where the asymptotic behavior of $V(z)$ in complex plane, as $|z| \rightarrow \infty$ and ${\rm Im}(z) \geq 0$, is of the form $V(z) \sim (C/z^{\delta}) \exp(i q z)$ with $q$ real and positive. Singularities of $V(z)$ at $|z| = \infty$, ${\rm Im}(z) \geq 0$ should be avoided, such as the case $V(z) \sim \cos(z) / z^{\delta}$, a potential which is not one-way reflectionless in spite it is holomorphic in the ${\rm Im}(z) \geq 0$ domain, but with a singularity at $z= \infty$.
\end{thebibliography}
\end{document}